\documentstyle[nolta,psfig]{article}
\begin{document}

\title{Localized Optimal Control of Spatiotemporal Chaos.}
\author{Roman O. Grigoriev$^1$, Sanjay G. Lall$^2$ and Geir E. Dullerud$^3$}
\affiliation{$1$ Condensed Matter Physics 114-36,
California Institute of Technology, Pasadena, CA 91125, USA\\
$2$ Control and Dynamical Systems 107-81,
California Institute of Technology, Pasadena, CA 91125, USA\\
$3$ Department of Applied Mathematics, University of Waterloo,
Waterloo, Ontario, Canada N2L 3G1}

\maketitle

\abstract
A linear output feedback control scheme is developed for a coupled map lattice
system. $H_\infty$ control theory is used to make the scheme local: both the
collection of information and the feedback are implemented through an array of
locally coupled control sites. Robustness properties of the control scheme are
discussed.
\endabstract

\section{Introduction}

Learning to tame spatiotemporal chaos in spatially extended nonlinear systems
is very attractive due to a large number of potential applications. Some of
these are continuous, such as turbulence \cite{lee}, plasma instabilities
\cite{pentek} and chemical reaction systems \cite{crowley}, some are discrete:
neural networks \cite{babloyantz} and distributed memory systems are only a few
examples. The main objective is usually to stabilize some suitable unstable
periodic orbit (UPO), or a group of orbits, embedded in the chaotic attractor
of the system.

Although spatially extended homogeneous systems could be treated as a special
case of the high-dimensional chaotic systems, some of the practical issues, that
arise in the control problem are quite specific and could be best handled by
taking into account the spatiotemporal structure of the system and the
controlled state in general and their symmetry properties in particular
\cite{self_pre}.   

In the present paper we will illustrate the control algorithm applying it to
the general coupled map lattice (CML), originally introduced by Kaneko
\cite{kaneko}:  
 \begin{equation}
 \label{eq_cml}
 z_i^{t+1}=f(z_i^t)+\epsilon(f(z_{i-1}^t)-2f(z_i^t)+f(z_{i+1}^t)),
 \end{equation}
and considered to be one of the simplest models, possessing the essential
properties of an extended spatiotemporally chaotic system.

There are many ways to achieve the stabilization of a non-chaotic trajectory.
However, the requirements imposed by different control algorithms and their
performance could vary widely. For instance, it was shown \cite{gang}, that a
number of UPOs of the CML (\ref{eq_cml}) could be stabilized with feedback
applied through a periodic array of controllers. Although limited knowledge of
the system state was required, the density of controllers had to be extremely
high for the control to work. Rearranging the controllers, one can
significantly reduce their density and improve the robustness characteristics
of the control scheme \cite{self_prl} at the expense of requiring additional
information about the system state. In the present paper we will show how the
CML can be controlled using low density of controllers and requiring very
limited information about the system state.

\section{The system}

Rewrite eq. (\ref{eq_cml}), adding to it the uncorrelated random noise
$\langle w^t_iw^{t'}_{i'}\rangle=\sigma^2\delta_{tt'}\delta_{ii'}$ and applying
control perturbations $u^t_k=G_k({\bf z}^t,{\bf z}^{t-1},\cdots)$ at sites
$i_k$, $k=1,\cdots,m$:
 \begin{eqnarray}
 \label{eq_cml_gen}
 z_i^{t+1}=\epsilon f(z_{i-1}^t)+(1-2\epsilon)f(z_i^t)
 +\epsilon f(z_{i+1}^t)\cr+w^t_i+\sum_k\delta_{ii_k}u^t_k,
 \end{eqnarray}
assuming, that the lattice is finite, $i=1,2,\cdots,n$, and periodic boundary
conditions $z_{i+n}^t=z_i^t$ are imposed.

Due to the translational symmetry of the CML (\ref{eq_cml}) additional
parameters can only enter the evolution equation through the nonlinear local
map function, which we choose as $f(z)=az(1-z)$, emphasizing, that the only
result affected by this particular choice is the set of existing periodic
trajectories. In particular, for any choice of $f(z)$, the homogeneity of the
system response to the perturbation of any internal parameter ($a$ and
$\epsilon$ in our case) makes it impossible to use either internal parameter
for control.

Linearizing equation (\ref{eq_cml_gen}) around the period-$\tau$ target UPO
$\hat{\bf z}^1,\hat{\bf z}^2,\cdots,\hat{\bf z}^\tau$ and denoting the
displacement ${\bf x}^t={\bf z}-\hat{\bf z}^t$, we obtain
 \begin{equation}
 \label{eq_cml_lin}
 {\bf x}^{t+1}=A^t{\bf x}^t+B^t_N{\bf w}^t+B^t{\bf u}^t,
 \end{equation}
where $A^t_{ij}=\partial_j z^{t+1}_i(\hat{\bf z}^t)$ is the Jacobian and the
matrices $B^t_N=I_{n\times n}$ and $B^t_{ij}=\sum_k\delta_{jk}\delta_{ii_k}$
specify the response of the system to the external noise $w^t_i$ and the
applied feedback $u^t_k$ (also called the {\em input}).

Finally, assume that only $q$ functions $\eta^t_i=H_i({\bf z}^t)$ (called 
the {\em output}) of the system state are accessible to measurement.
Denoting $C^t_{ij}=\partial_j H_i(\hat{\bf z}^t)$ we obtain for the
linearized output:
 \begin{equation}
 \label{eq_output}
 y^t_i=\eta^t_i-H_i(\hat{\bf z}^t)=C^t_{ij} x^t_j.
 \end{equation}

\section{The control scheme}

The algorithm presented below allows one to determine whether the feedback
${\bf u}^t$ stabilizing the chosen UPO can be obtained as a function of the
output ${\bf y}^t$, and determines the solution, which minimizes the noise
amplification factor or {\em induced-power-norm}
 \begin{equation}
 \gamma=\max_{\|{\bf w}\|_P<\infty}\frac{\|{\bf z}\|_P}{\|{\bf w}\|_P},
 \end{equation}
where the $r$-dimensional performance vector
 \begin{equation}
 \label{eq_perform}
 {\bf z}^t=C_N^t{\bf x}^t+D_N^t{\bf u}^t
 \end{equation}
gives the deviation of the system from the target state,
and the power norm is defined as\vskip -0.7mm
 \begin{equation}
 \|{\bf z}\|_P=\left[\lim_{T\rightarrow\infty}\frac{1}{T}
 \sum_{t=1}^T|{\bf z}^t|^2\right]^{1/2}.
 \end{equation}

The solution to the time-periodic {\em output feedback} problem 
(\ref{eq_cml_lin},\ref{eq_output},\ref{eq_perform}) can be obtained using the
generalization of the results of $H_\infty$ control theory \cite{doyle} for
linear time invariant (LTI) systems. In particular, Dullerud and Lall have
shown \cite{lall}, that if a locally stabilizing linear feedback ${\bf u}^t$
exists, it could be written as
 \begin{eqnarray}
 \label{eq_control}
 {\bf v}^{t+1}&=&A_C^t{\bf v}^t+B_C^t{\bf y}^t\cr
 {\bf u}^t&=&C_C^t{\bf v}^t+D_C^t{\bf y}^t,
 \end{eqnarray}
where $A_C^t,B_C^t,C_C^t$ and $D_C^t$ are matrices with the same periodicity
$\tau$ as the target orbit $\hat{\bf z}^t$, and ${\bf v}^t$ is the
$p$-dimensional internal state of the controllers. The standard {\em state
feedback} law ${\bf u}^t=K^t{\bf x}^t$ used in \cite{self_prl}, is seen to be
just a special case of this general setup.

Construct constant block diagonal matrices $A$,$B$,$C$, $B_N$,$C_N$ and
$D_N$ according to the following rule:
 \begin{equation}
 Q=\left[\matrix{Q^1& \cdots& 0\cr \vdots& \ddots& \vdots\cr 0& \cdots& Q^\tau}
 \right].
 \end{equation}
For $\tau>1$ define a $\tau n\times\tau n$ cyclic shift matrix
 \begin{equation}
 Z=\left[\matrix{0_{n\times n} & \cdots & 0_{n\times n} & I_{n\times n}\cr
         I_{n\times n} & \cdots & 0_{n\times n} & 0_{n\times n}\cr
         \vdots & \ddots & \vdots & \vdots\cr
         0_{n\times n} & \cdots & I_{n\times n} & 0_{n\times n}} \right].
 \end{equation} 
In the time-invariant case ($\tau=1$) set $Z=I_{n\times n}$. Also introduce the
notations $Q>0$ for positive definite, $Q\ge0$ for semi-positive definite
matrices and $Q^\dagger$ for the transpose of $Q$. 

It can be shown \cite{lall}, that a stabilizing solution
(\ref{eq_control}) with $p\ge n$ such that $\gamma<1$ for the
system (\ref{eq_cml_lin}-\ref{eq_perform}) exists, if and only if
there exist block-diagonal matrices $R>0$ and $S>0$, satisfying
 \begin{equation}
 \label{eq_condition}
 \left[\matrix{R&I\cr I&S}\right]\ge 0\,\quad
 O_S^\dagger P_S O_S<0\,\quad
 O_R^\dagger P_R O_R<0
 \end{equation}
where $P_R,P_S,O_R$ and $O_S$ are given by
 \begin{eqnarray}
 \hskip -12mm P_S=\left[\matrix{
     A^\dagger Z^\dagger SZA-S &  A^\dagger Z^\dagger SZB_N & 
                                     C_N^\dagger\cr
      B_N^\dagger Z^\dagger SZA & 
                    B_N^\dagger Z^\dagger SZB_N-I & 0\cr
     C_N & 0 & -I}\right]\hskip -7mm\cr
 \hskip -2mm
 P_R=\left[\matrix{ARA^\dagger-Z^\dagger RZ & ARC_N^\dagger &  B_N\cr
                   C_N RA^\dagger & C_N^\dagger RC_N-I & 0\cr
                    B_N^\dagger & 0 & -I}\right]\quad\hskip -5mm\cr
 O_S=\left[\matrix{N_S & 0\cr 0 & I}\right],\qquad
 O_R=\left[\matrix{N_R & 0\cr 0 & I}\right]\qquad
 \end{eqnarray}
and the unitary matrices $N_R$ and $N_S$ satisfy
 \begin{eqnarray}
 {\rm Im}\, N_R={\rm ker}\left[\matrix{B^\dagger & D_N^\dagger}\right]\;\cr
 {\rm Im}\, N_S={\rm ker}\left[\matrix{C & 0_{p\times n}}\right].
 \end{eqnarray}
 
To minimize $\gamma$, rescale $C_N^t$ and $D_N^t$, such that the above
condition tests for $\gamma<\gamma_0$ instead of $\gamma<1$ and decrease
$\gamma_0$ until the test fails; standard software exists to do this. If there
is \emph{any} linear stabilizing controller, we can therefore find it using
this algorithm.

If $R={\rm diag}(R^1,\dots,R^\tau)$ and $S={\rm diag}(S^1,\dots,S^\tau)$ are
determined, one can find the matrices in (\ref{eq_control}) using the
following procedure. First, construct nonsingular matrices $M_t$ and $N_t$,
such that
 \begin{equation}
 M_tN_t^\dagger=I-R^tS^t.
 \end{equation}
Determine the matrix $X_t$ as the unique solution of
 \begin{equation}
 \left[\matrix{S^t&I \cr N_t^\dagger&0}\right]
 =X_t\left[\matrix{I&R^t \cr 0&M_t^\dagger}\right].
 \end{equation}
Next, define the matrices
 \begin{eqnarray}
 \tilde{A}_t=\left[\matrix{A^t&0_{n\times n}\cr
                   0_{n\times n}&0_{n\times n}}\right]\quad\;
 \tilde{B}_t=\left[\matrix{B_N^t\cr 0_{n\times n}}\right]\qquad\quad\;\cr
 \tilde{C}_t=\left[\matrix{C_N^t& 0_{n\times n}}\right]\qquad\,\,
 \hat{B}_t=\left[\matrix{0_{n\times n}&I_{n\times n}\cr
                   B^t&0_{m\times n}}\right]\cr
 \hat{C}_t=\left[\matrix{0_{n\times n}&I_{n\times n}\cr
                   C^t& 0_{p\times n}}\right]\quad\;
 \hat{D}_t=\left[\matrix{0_{r\times n} & D_N^t}\right]\quad\;
 \end{eqnarray}
and then define
 \begin{eqnarray}
 H_t=\left[\matrix{-X_{t+1}^{-1} &\tilde{A}_t &\tilde{B}_t &0_{2n\times r}\cr
   \tilde{A}^\dagger_t & -X_t & 0_{2n\times n} & \tilde{C}_t\cr
   \tilde{B}_t^\dagger & 0_{n\times 2n} & -I_{n\times n} & 0_{n\times r}\cr
   0_{r\times 2n} & \tilde{C}_t & 0_{r\times n} & -I_{r\times r}
                  }\right]\;\cr
 Q_t=\left[\matrix{0_{n+p\times 2n} & \hat{C}_t & 0_{n+p\times n} &
                   0_{n+p\times r}}\right]\quad\,\cr
 P_t=\left[\matrix{\hat{B}_t^\dagger & 0_{n+m\times 2n} & 0_{n+m\times n} &
                   \hat{D}_t^\dagger}\right]\qquad
 \end{eqnarray}
Finally, the matrices $A_C^t,B_C^t,C_C^t$ and $D_C^t$ are extracted from the
solution
 \begin{equation}
 J_t=\left[\matrix{A_C^t&B_C^t \cr C_C^t&D_C^t}\right]
 \end{equation}
to the linear matrix inequality
 \begin{equation}
 \label{eq_jhpq}
 H_t+Q_t^\dagger J_t^\dagger P_t+P_t^\dagger J_tQ_t<0.
 \end{equation}
 
Linear matrix inequalities (LMI) like (\ref{eq_condition}) and (\ref{eq_jhpq})
can be conveniently solved using the tools of convex optimization theory. The
big practical advantage of this technique is the guaranteed convergence.

\begin{figure}
\centering
\mbox{\psfig{figure=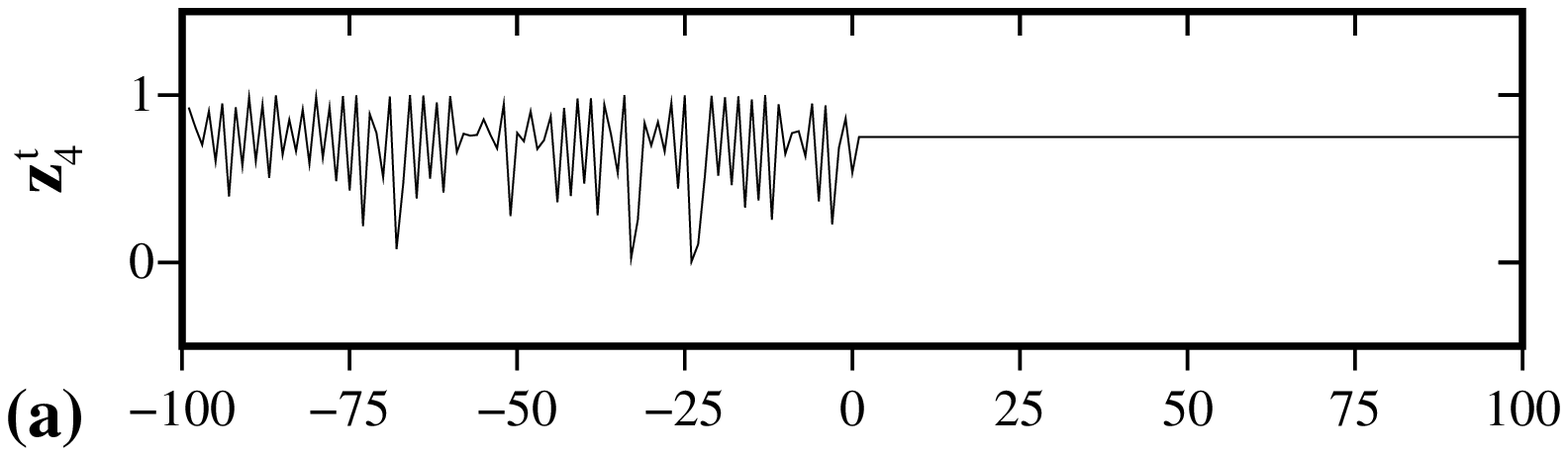,width=3.2in}}
\mbox{\psfig{figure=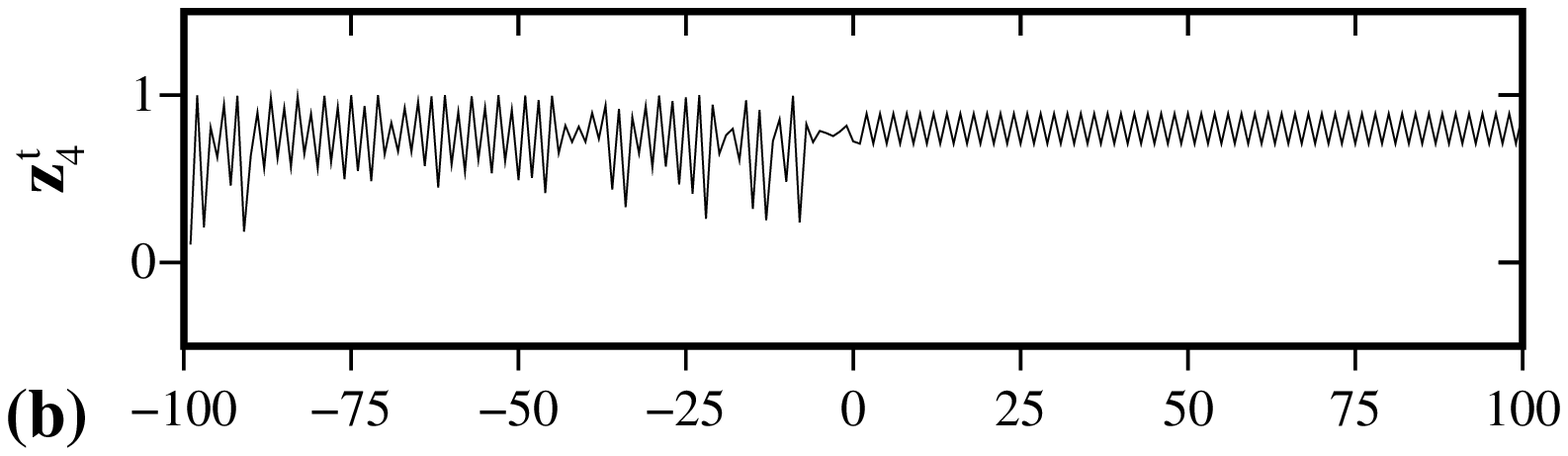,width=3.2in}}
\mbox{\psfig{figure=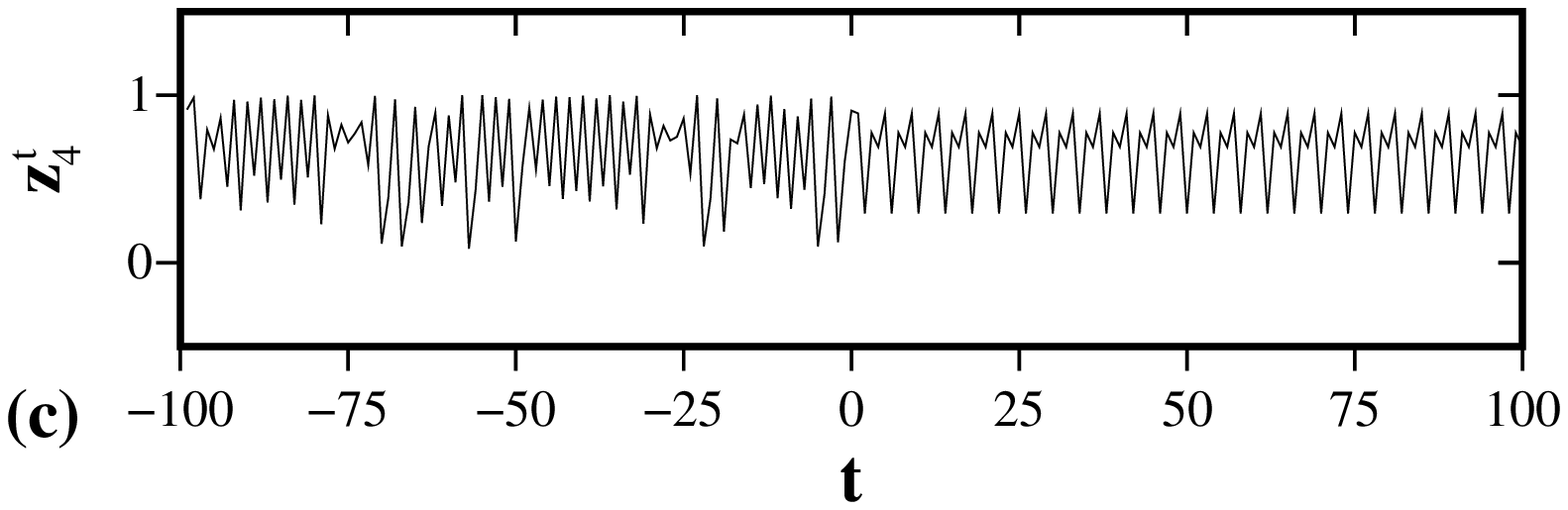,width=3.2in}}
\caption{ Stabilization of unstable periodic orbits of the noisy system: (a)
S1T1, (b) S8T2 and (c) S8T4. 4th lattice variable $z^t_4$ is plotted. The time of
capture is taken to be $t=0$. The noise strength is $\sigma=10^{-6}$.}
\label{fig_upos}
\end{figure}

\section{Control of large lattices}

Although, using the above algorithm, we can in principle obtain the
stabilizing feedback (\ref{eq_control}) for a system (\ref{eq_cml_gen}) of
arbitrary size, solving matrix inequalities involving large matrices requires
considerable computational resources.

This problem could be avoided using distributed control approach. The idea is
to subdivide the complete system into a number of weakly interacting
subsystems, and learn to control each of the subsystems independently,
neglecting interactions with other subsystems. Finally, the control can be
adjusted to correct for interactions by introducing coupling between formerly
independent controllers. 

Since the coupling in our model is local, we can partition the whole lattice
into a number of identical subdomains of length $n_p\ll n$, each interacting
with two adjacent subdomains. The original problem is thus reduced to the
problem of controlling an isolated subdomain of limited length $n_p$ (we drop
the index below). We impose periodic boundary conditions on each subdomain to
allow the existence of unstable orbits periodic in space as well as time.

The symmetry properties of the CML (\ref{eq_cml}) determine \cite{self_pre},
that the minimal number of controllers required is two. Placing them at the
boundaries of the subdomain allows one to change the boundary condition at
will, as well as correct for interactions between adjacent subdomains, by
adding appropriate perturbations to the feedback \cite{self_prl}. This defines
the matrix
 \begin{equation}
 B^t_{ij}=\delta_{i1}\delta_{j1}+\delta_{in}\delta_{j2}.
 \end{equation}

In order to calculate these perturbations we will have to introduce coupling
between controllers of adjacent subdomains. Specifically, we will need to
exchange the information about the state of the system in the neighborhood of
the boundaries (and therefore controllers), i.e. at least the variables $x^t_1$
and $x^t_n$ should be measurable. This defines the minimal
realization of the matrix $C^t_{ij}=B^t_{ji}$, $q=m=2$, which we use below.

\section{Comparison of $H_2$ and $H_\infty$ approaches}

In order to compare the results of the proposed approach with those, obtained
using linear quadratic ($H_2$) theory for the state feedback \cite{self_prl},
we select a similar optimization criterion. Specifically, we take
 \begin{equation}
 C_N=\left[\matrix{I_{n\times n}\cr 0_{m\times n}}\right]\qquad
 D_N=\left[\matrix{0_{n\times m}\cr I_{m\times m}}\right],
 \end{equation}
such that $r=n+m$, ${\bf z}^t=\{{\bf x}^t;{\bf u}^t\}$ and, consequently,
 \begin{equation}
 ||{\bf z}||^2_P=\lim_{T\rightarrow\infty}\frac{1}{T}\sum_{t=1}^T\left(
 {{\bf x}^t}^\dagger{\bf x}^t+{{\bf u}^t}^\dagger{\bf u}^t\right).
 \end{equation}

\begin{figure}
\centering
\mbox{\psfig{figure=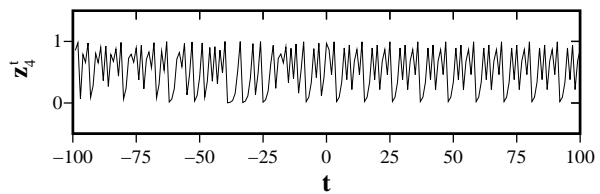,width=3.2in}}
\caption{ Stabilization of the homogeneous time-period-11 (S1T11) UPO. The
time of capture is taken to be $t=0$. The noise strength is $\sigma=10^{-6}$.}
\label{fig_long}
\end{figure}

We demonstrate the $H_\infty$ approach by stabilizing a number of UPOs of the
noisy CML (\ref{eq_cml_gen}) with $n=8$ sites, $a=4.0$ and $\epsilon=0.33$. The
feedback (\ref{eq_control}) is calculated using the algorithm outlined above.
Figure \ref{fig_upos} shows the process of capturing and controlling the steady
homogeneous state (S1T1), the time-period-2 space-period-8 (S8T2), and the
time-period-4 space-period-8 (S8T4) orbits.

The real power of the $H_\infty$ approach, however, can be full appreciated
only in application to orbits of very high periodicity, where the accurate
treatment of the effects of noise is of ultimate importance. Any method based
on the reduction of periodic trajectories to steady states will fail for orbits
of sufficiently long period. The $H_\infty$ approach does not suffer from this
limitation. Indeed, we have observed stabilization of a number of periodic
orbits with period $\tau>10$. One such example is presented in Fig.
\ref{fig_long}. 

Noise limits our ability to control arbitrarily large systems with local
interactions, using just two controllers. Rather simple arguments show
\cite{self_prl}, that the size of the largest system, that could be stabilized
in the presence of random perturbations ${\bf w}^t$, could be estimated using
the controllability condition, if {\em complete} information about the state of
the system is available. So, for a steady uniform state one obtains
 \begin{equation}
 \label{eq_max_2}
 n_2(\sigma)=
 \cases{-\lambda_{max}^{-1}\log(\sigma),&$\epsilon>0.5$\cr
 2(\log(\epsilon)-\lambda_{max})^{-1}\log(\sigma),&$\epsilon<0.5$,}
% \cases{-\frac{\log(\sigma)}{\lambda_{max}}, & if $\;0.5<\epsilon<1$\cr
% \frac{2\log(\sigma)}{\log(\epsilon)-\lambda_{max}}, & if $\;0<\epsilon<0.5$,}
 \end{equation}
where $\lambda_{max}$ is the maximal Lyapunov exponent.

\begin{figure}
\centering
\mbox{
\psfig{figure=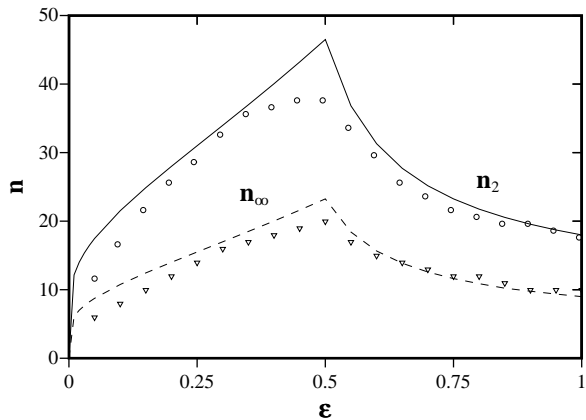,width=3.2in}}
\caption{ The largest length of the lattice, which can be controlled with two
pinning sites: the triangles represent the length obtained using $H_\infty$
control, the circles show the data, obtained using $H_2$ control in Ref.
\cite{self_prl}, and the curves show theoretical estimates (\ref{eq_max_2})
and (\ref{eq_max_inf}). The noise strength is $\sigma=10^{-14}$ and $a=4.0$. }
 \label{fig_length}
\end{figure}

If however only {\em partial} information about the state of the system is
available, additional requirements appear. Any control algorithm utilizing
output feedback essentially consists of two major stages: {\em observation} and
{\em control}. During the first stage information about the system is collected
and processed to recreate the state of the system. During the second stage,
control perturbations are applied to bring the system to the desired state. As
a result the control scheme should be able to tolerate uncertainties introduced
during both the control and the observation stage.

The estimate (\ref{eq_max_2}) reflects the requirements imposed by the control
stage. Additional requirements, introduced by the observation stage can be
similarly estimated using the {\em observability} condition,
 \begin{equation}
 {\rm rank}\left[\matrix{C^\dagger\quad A^\dagger C^\dagger\quad \cdots\quad
 {A^\dagger}^{n-1} C^\dagger}\right]=n,
 \end{equation}
which determines whether the state of the system can be extracted from the
observed data (\ref{eq_output}), and for $C=B^\dagger$ coincides with the
controllability condition.

Careful consideration shows, that the addition of the observation stage
effectively doubles both the length of the control cycle and the length of the
lattice. As a result, the maximal length of the system, that can be
successfully stabilized using $H_\infty$ control is halved:
 \begin{equation}
 \label{eq_max_inf}
 n_\infty(\sigma)=n_2(\sigma)/2.
 \end{equation}

The maximal length $n_\infty(\sigma)$ can be obtained numerically by choosing
the fixed point as the initial condition and monitoring the evolution of the
system in the presence of noise under control (\ref{eq_control}). The results
are presented in Fig. \ref{fig_length}. One can see that the estimate
(\ref{eq_max_inf}) approximates the actual results rather well.

This work was partially supported by the NSF through grant no. DMR-9013984.

\end{document}